\documentclass[aps,prl,superscriptaddress,twocolumn,amsmath,amssymb]{revtex4}

\usepackage{color}
\usepackage{graphicx}
\usepackage{bm}
\definecolor{Blue}{rgb}{0.00, 0.00, 1.00}
\definecolor{Red}{rgb}{1.00, 0.00, 0.00}

\begin{document}

\title{Novel mid-infrared plasmonic properties of bilayer graphene}

\author{Tony Low}
\email{tonyaslow@gmail.com}
\affiliation{IBM T.J. Watson Research Center, 1101 Kitchawan Rd, Yorktown Heights, NY 10598, USA}
\author{Francisco Guinea}
\affiliation{Instituto de Ciencia de Materiales de Madrid. CSIC. Sor Juana In\'es de la Cruz 3. 28049 Madrid, Spain}
\author{Hugen Yan}
\affiliation{IBM T.J. Watson Research Center, 1101 Kitchawan Rd, Yorktown Heights, NY 10598, USA}
\author{Fengnian Xia}
\affiliation{IBM T.J. Watson Research Center, 1101 Kitchawan Rd, Yorktown Heights, NY 10598, USA}
\affiliation{Yale University, 15 Prospect Street, New Haven, CT 06511, USA}
\author{Phaedon Avouris}
\affiliation{IBM T.J. Watson Research Center, 1101 Kitchawan Rd, Yorktown Heights, NY 10598, USA}
\date{\today}
\begin{abstract}

We study the mid-infrared plasmonic response in Bernal-stacked bilayer graphene.
Unlike its monolayer counterpart, bilayer graphene accommodates optically active
phonon modes and a resonant interband transition at infrared frequencies.
They strongly modifies the plasmonic properties of bilayer graphene,
leading to Fano-type resonances, giant plasmonic enhancement of infrared phonon absorption,
narrow window of optical transparency, and a new plasmonic mode at higher energy than the classical plasmon.
\end{abstract}
\maketitle

Plasmonics\cite{maier2007plasmonics} is an important subfield of photonics that
deals with the excitation, manipulation, and utilization of plasmons-polaritons\cite{pines1999elementary}.
It is a key element of nanophotonics\cite{gramotnev2010plasmonics},
metamaterials with novel electromagnetic phenomena\cite{shalaev2007optical,luk2010fano}
and also has potential applications in biosensing\cite{kabashin2009plasmonic}.
Recently, graphene has emerged as a promising platform for plasmonics\cite{grigorenko2012graphene}.
It has many desirable properties such as gate-tunability, extreme light confinement, long plasmon lifetime,
and plasmonic resonances in the terahertz to mid-infrared (IR) regime\cite{JSB11,KCG11,HSS07,WSSG06,NGGM12,nikitin2011edge}.
Spatially resolved propagating plasmons has been observed with scanning near-field optical microscope\cite{FRABM12,CBATH12}.
Tunable plasmon resonances in the terahertz\cite{JGHGM11} to
IR\cite{YLC12,YLZW13} has been observed in graphene micro- and nano-ribbons, and the relative
damping pathways have also been studied\cite{YLZW13}.
Identified applications for graphene plasmonics range from notch filters\cite{YLC12}, polarizers and modulators\cite{JGHGM11,YLC12,YLZW13} to
beam reflectarrays\cite{carrasco2013tunable}, biosensing\cite{wu2010highly}
and IR photodetectors\cite{freitag2013photocurrent} via bolometric effect\cite{freitag2012photoconductivity}.

In this paper, we discuss why Bernal AB-stacked bilayer graphene is important and interesting in its own right as a plasmonic material.
Apart from a few theoretical studies of plasmons in bilayer graphene\cite{sensarma2010dynamic,gamayun2011dynamical,gorbar2010dynamics,borghi2009dynamical,kusminskiy2009electron,hwang2010plasmon},
there is still no experimental studies of bilayer graphene plasmonics.
First indication that the plasmonic response in bilayer graphene might be very different
than that of monolayer is its two prominent IR structures in its optical conductivity.
IR optical measurements of bilayer graphene reveal a phonon peak at $\hbar\omega$$\,\approx\,$$0.2\,$eV,
with a strong dependence of peak intensity and Fano-type lineshape on the applied gate voltage\cite{tang2009tunable,kuzmenko2009gate}.
The interlayer coupling in bilayer graphene also results in
two nested bands, which presents a set of doping dependent IR features\cite{nilsson2006electronic,abergel2007optical,NC08}.
This interband transitions between the two nested bands
produced a conductivity peak at $\hbar\omega$$\,\approx\,$$0.4\,$eV in optical IR measurements\cite{wang2008gate,kuzmenko2009infrared,li2009band}. The impact of these IR structures on the bilayer
plasmonic response has not been studied.
We found several novel plasmonic effects in bilayer graphene:
(i) giant plasmonic enhancement of infrared phonon absorption,
(ii) an extremely narrow optical transparency window, and (iii) a new plasmonic mode at higher energy than the classical plasmon.

\begin{figure}[t]
\centering
\scalebox{0.45}[0.45]{\includegraphics*[viewport=90 265 500 590]{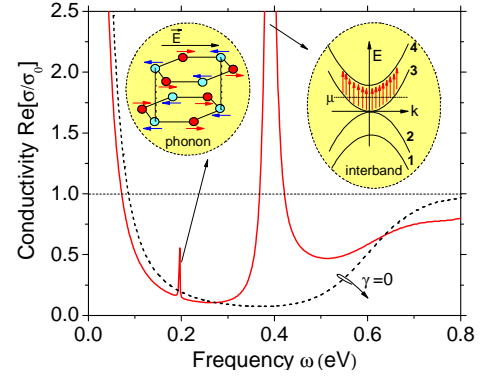}}
\caption{
Real part of bulk bilayer graphene conductivity (solid line) computed at $T=300\,$K
at chemical potential of $\mu=0.3\,$eV, constant damping of $\eta=10\,$meV, zero gap (i.e. $\Delta=0\,$eV)
and $q=0$.
This is compared with the case where $\gamma_1=0\,$eV (dashed line).
$\sigma_0$ is universal conductivity of $e^2/2\hbar$.
}
\label{figure1}
\end{figure}

Bilayer graphene arranged in the Bernal AB stacking order is considered,
with basis atoms $A_1$, $B_1$ and  $A_2$, $B_2$ in the top and bottom layers respectively.
The intralayer coupling is $\gamma_0\approx 3\,$eV  and the interlayer coupling between
$A_2$ and $B_1$ is  $\gamma_1\approx 0.39\,$eV,
an average of values reported in optical IR and photoemission measurements\cite{kuzmenko2009infrared,wang2008gate,li2009band,ohta2006controlling,zhou2008origin}.
We work within the $4\times 4$ atomic $p_z$ orbitals basis, i.e.
$a^{\dagger}_{1\bold{k}},b^{\dagger}_{1\bold{k}},a^{\dagger}_{2\bold{k}},b^{\dagger}_{2\bold{k}}$,
where $a^{\dagger}_{i}$ and $b^{\dagger}_{i}$ are creation operators for the $i^{th}$
layer on the $A/B$ sublattices.
Within this basis, the Hamiltonian near the $\bold{K}$ point can be written as:
${\cal H}_{k}=v_f \pi_{+}I\otimes\sigma_{-}+v_f \pi_{-}I\otimes\sigma_{+}+\tfrac{\Delta}{2}\sigma_z\otimes I+\gamma_1/2[\sigma_x\otimes\sigma_x+\sigma_y\otimes\sigma_y]$,
where $\sigma_i$ and $I$ are the Pauli and identity matrices respectively.
We defined $\sigma_{\pm}\equiv\tfrac{1}{2}(\sigma_x \pm i \sigma_y)$
and $\pi_{\pm}\equiv\hbar(k_x\pm ik_y)$. Here, \textcolor{black}{$v_f$ is the in-plane velocity\cite{mccann06}} 
and $\Delta$ is the electrostatic potential
difference between the two layers. Expressions for non-interacting ground state
electronic bands $\xi_{n}(\bold{k})$ ($n=1-4$, see inset of Fig.\,\ref{figure1}) and wavefunctions $\Phi_n(\bold{k})$
are obtained by diagonalizing ${\cal H}_{k}$, see
Suppl. Info.

We consider coupling of long wavelength longitudinal/transverse optical (LO/TO) phonons near $\Gamma$ point
with the graphene plasmons.
Relative displacement of the two sublattice in the \textcolor{black}{top layer ($T$)}
is given by,
\begin{align}
\bold{u}_T(\bold{r}) = \sqrt{\frac{\hbar}{2\rho_m \omega_{op}{\cal A}}}\sum_{\bold{p}\lambda} (\hat{b}_{\bold{p}}+\hat{b}_{\mbox{-}\bold{p}}^{\dagger})
\bold{e}_{\lambda}(\bold{p})  e^{i\bold{p}\cdot\bold{r}}
\end{align}
where ${\cal A}$ is the area of the unit cell, $\rho_m$ is the mass density of graphene, $\bold{p}=(p_x,p_y)$ is
the phonon wavevector, $\lambda$ denotes the LO/TO modes where
$\hat{b}_{\bold{p}\lambda}^{\dagger}$ is its
creation operators,
$\bold{e}_{\lambda}(\bold{p})$ are the polarization vectors given by
$\bold{e}_{LO}(\bold{p})=i(\mbox{cos}\varphi,\mbox{sin}\varphi)$ and
$\bold{e}_{TO}(\bold{p})=i(-\mbox{sin}\varphi,\mbox{cos}\varphi)$ where
$\varphi=\mbox{tan}^{-1}(p_y/p_x)$.
Due to the two graphene layers, there are two possible vibrational modes i.e. symmetric
($\bold{u}_B(\bold{r})$=$\bold{u}_T(\bold{r})$) and antisymmetric
($\bold{u}_B(\bold{r})$=$-\bold{u}_T(\bold{r})$), \textcolor{black}{where subscript $B$ denotes bottom layer.}
Hence, the electron-phonon coupling at the $\bold{K}$ valley
for bilayer graphene is given by\cite{ando07,NG07},
\begin{align}
H_{e-op}(\bold{r})=-\sqrt{2}\frac{2\beta\hbar v_F}{3 a^2} \boldsymbol{\sigma}^{\pm}\times \bold{u}(\bold{r})
\end{align}
with $a\approx 1.4$ \AA \, is the C-C distance, $\sigma_j^+$=$I\sigma_j$, $\sigma_j^{-}$=$\sigma_z\sigma_j$
and $\beta$=$-\partial \mbox{ln} \gamma_0/\partial a$ is a dimensionless
parameter related to the deformation potential.
Without loss of generality,
we take the electric field polarization to be along $y$ and $\varphi=0$.
Since only lattice vibration along $y$ can couple to light,
we consider only the TO lattice mode. As a result, we can write the
electron-phonon interaction for the $v$ mode in the following form,
\begin{align}
{\cal H}'_v = \frac{1}{\sqrt{{\cal A}}}\sum_{\bold{k}} \hat{a}_{\bold{k+p}}^{\dagger} {\cal V}_v(\bold{p})
\hat{a}_{\bold{k}} e^{i\bold{p}\cdot\bold{r}}(\hat{b}_{\bold{p}}+\hat{b}_{\bold{p}}^{\dagger})
\end{align}
where  $v=S,A$ denotes the symmetric and antisymmetric modes, with
${\cal V}_S(\bold{p}\rightarrow 0)=ig I\sigma_x$ and ${\cal V}_A(\bold{p}\rightarrow 0)=ig \sigma_z\sigma_x$, where,
\begin{eqnarray}
g\equiv \frac{\beta\hbar v_F}{L^2}\sqrt{\frac{\hbar}{2\rho_m\omega_{op} }}\approx 0.3\,eV\AA^{-1},
\end{eqnarray}
since  $\beta\approx 2$ and $\hbar\omega_{op} \approx 0.2\,$eV\cite{ando07}.

The plasmonic response of bilayer graphene can be obtained from its
dielectric function given by,
\begin{align}
\epsilon_T^{rpa}(q,\omega)=\kappa-v_{c}\Pi^0_{\rho,\rho}(q,\omega)-v_{c}\frac{q^2}{\omega^2}\delta\Pi_{j,j}(q,\omega),
\end{align}
at arbitrary wave-vector $q$ and frequency $\omega$.
$v_c=e^2/2q\epsilon_0$ is the $2D$ Coulomb interaction \textcolor{black}{and $\kappa$ is the effective dielectric constant of the environment.}
$\Pi_{\rho,\rho}^0(q,\omega)$ is the non-interacting part (i.e. the pair bubble diagram) of
the charge-charge correlation function given by\cite{WSSG06,HSS07},
\begin{eqnarray}
\nonumber
\Pi_{\rho,\rho}^0(q,\omega)=-\frac{g_s g_v}{(2\pi)^2}\sum_{nn'}\int d\bold{k} \times\\
\frac{n_F(\xi_{n}(\bold{k}))-n_F(\xi_{n'}(\bold{k+q}))}{\xi_{n}(\bold{k})-\xi_{n'}(\bold{k+q})+\hbar\omega+i\hbar/\tau_e}\left|F_{nn'}(\bold{k},\bold{q})\right|^2
\end{eqnarray}
where $n_F$ is the Fermi-Dirac distribution function,
\textcolor{black}{$g_s$ and $g_v$ are the spin/valley degeneracy,
}$F_{nn'}(\bold{k},\bold{q})$=$\left\langle \Phi_n(\bold{k}) \right.\left| \Phi_{n'}(\bold{k+q})\right\rangle$ is the band overlap,
and $\tau_e$ is the electron lifetime, where we assumed a typical experimental value of
$\eta\equiv\hbar/\tau_e\approx 10\,$meV\cite{YLZW13}.

The effect  of electron-phonon interaction is included within  $\delta\Pi_{j,j}(q,\omega)$,
\textcolor{black}{where subscript $j$ denotes the current operator.}
Here, we employ a model for $\delta\Pi_{j,j}(q,\omega)$ which is consistent with the various
electron-phonon selection
rules for the symmetric/antisymmetric modes and Fano effect observed in optical spectroscopy
experiments for bilayer graphene.
The detailed implementation follows a formalism known as the charged-phonon theory\cite{rice1992charged,CBK10,CBMK12},
\begin{align}
\delta\Pi_{j,j}(q,\omega)=\sum_{vv'}\Gamma_{j,v}(q,\omega){\cal D}_{vv'}(\omega)\Gamma_{v'^{\dagger},j}(q,\omega)
\end{align}
where
\begin{eqnarray}
\nonumber
\Gamma_{j,v}(q,\omega) = -\frac{g_s g_v}{(2\pi)^2}\sum_{nn'}\int d\bold{k} \times\\
\frac{n_F(\xi_{n}(\bold{k}))-n_F(\xi_{n}(\bold{k+q}))}{\xi_{n}(\bold{k})-\xi_{n'}(\bold{k+q})+\hbar\omega+i\hbar/\tau_e}
[{\cal J}]_{nn'} [{\cal V}_v]_{n'n}
\end{eqnarray}
where $\left[{\cal J}\right]_{nn'} = \left\langle \Phi_n(\bold{k}) \right|{\cal J}  \left| \Phi_{n'}(\bold{k+q})\right\rangle$
and $\left[{\cal V}_v\right]_{nn'} = \left\langle \Phi_n(\bold{k}) \right|{\cal V}_v  \left| \Phi_{n'}(\bold{k+q})\right\rangle$
with $v=A,S$ and the current operator defined as
${\cal J}\equiv v_F I\sigma_y$ with the direction of the electric field.
${\cal D}$ is the phonon Green's function,
\begin{align}
[{\cal D}^{-1}(\omega)]_{vv'} = \delta_{vv'}[{\cal D}_0^{-1}(\omega)]-\Gamma_{v^{\dagger},v'}(\omega)
\end{align}
where ${\cal D}_0=2\omega_{op}/\hbar((\omega+i/\tau_{op})^2-\omega_{op}^2)$ is the
free phonon Green's function and $\tau_{op}$ describes the phonon lifetime.
In this calculation, we assumed $\tau_{op}\approx 1\,$ps\cite{bonini2007phonon}.

Fig.\ref{figure1} shows the optical conductivity of bilayer graphene calculated
from the relation\cite{CBK10},
\begin{align}
\sigma(q,\omega) = \underbrace{i\frac{e^2\omega}{q^2}\Pi^0_{\rho,\rho}(q,\omega)}_{\bar{\sigma}} + \underbrace{i\frac{e^2}{\omega}\delta\Pi_{j,j}(q,\omega)}_{\delta\sigma}
\end{align}
The calculation assumes $T=300\,$K, chemical potential of $\mu=0.3\,$eV and $\Delta=0\,$eV.
$\bar{\sigma}$ is the non-interacting optical conductivity, which
accounts for a Drude peak at $\omega=0$ and a universal conductivity of $e^2/2\hbar$.
The peak conductivity at $\hbar\omega=\gamma$ is due to interband transitions between
two perfectly nested bands, e.g. $\xi_3$ and $\xi_4$, separated in energy by $\gamma$, see inset.
These conductivity peaks at $\omega=0$ and $\hbar\omega=\gamma$ are phenomenologically broadened by
$\omega\rightarrow\omega+i/\tau_e$ in the model.
$\delta\sigma$ accounts for the electronic interaction with the IR phonons modes ($v=A,S$),
and agrees well with experimentally measured optical spectra of bilayer graphene\cite{CBK10}.
In our zero gap case, only the $A$ mode (asymmetric mode)
is IR active\cite{CBK10}, see inset of Fig.\ref{figure1}.
This mode is responsible for the sharp resonance feature at $\omega=\omega_{op}$.

\begin{figure}[t]
\centering
\scalebox{0.45}[0.45]{\includegraphics*[viewport=90 80 500 730]{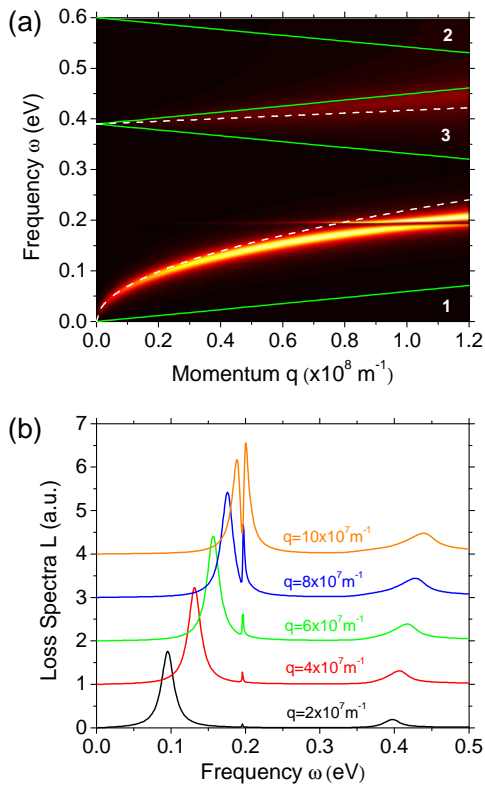}}
\caption{
$\bold{(a)}$ shows the RPA electron loss function $L(q,\omega)$ for bilayer graphene
computed at $T=300\,$K
at chemical potential of $\mu=0.3\,$eV, constant damping of $\eta=10\,$meV,
zero energy gap (i.e. $\Delta=0\,$eV) and a background dielectric constant of $\kappa=2.5$. Green lines are boundaries for the Landau damped regions.
Spectra at different plasmon momenta $q$ are plotted in $\bold{(b)}$.
}
\label{figure2}
\end{figure}


Longitudinal collective plasmonic dispersion
is obtained by looking for the zeros in
the real part of the dynamical dielectric function i.e. $\mbox{Re}[\epsilon_T^{rpa}(q,\omega)]=0$.
For bilayer graphene, there are three solutions\cite{sensarma2010dynamic,G11};
a `classical' plasmon with $\sqrt{q}$ behavior, an acoustic plasmon with $\propto q$ behavior
and a high energy $\gamma$-plasmon residing near the interband resonance $\gamma$.
Only the former has been found to be fully coherent, whose dispersion in the
long wavelength limit can be shown to follow,
\begin{align}
\omega_{pl}(q)=\frac{1}{\hbar}\sqrt{\frac{qe^2g}{4\pi\epsilon_0\kappa}\sum_j \frac{n_j(\mu)}{D_j(\mu)}}
\label{om_pl}
\end{align}
where $g=4$ is the degeneracy factor, $n_j(\mu)$ and $D_j(\mu)$
is the carrier density and density-of-states of the $j$-th band respectively.
On the other hand, the other two solutions are overdamped.
The acoustic plasmon lies in the intraband continuum and is
always overdamped with insignificant spectral weight\cite{sensarma2010dynamic,G11}.
Under typical conditions, the high energy $\gamma$-plasmon is also overdamped,
lying in the interband continuum (i.e. $\xi_1,\xi_2\rightarrow\xi_3,\xi_4$ transitions) when $2\mu<\gamma$ and
the low-energy interband continuum (i.e. $\xi_1\rightarrow\xi_2$ or $\xi_3\rightarrow\xi_4$ transitions) when $2\mu>\gamma$.
We show later that, under certain conditions, this mode can become fully coherent.

Electron loss function, defined as the imaginary part of the
inverse dielectric function i.e. $L(q,\omega)=[\epsilon_{T}^{rpa}(q,\omega)]^{-1}$,
is a quantity that can be probed in various spectroscopy experiments\cite{YLZW13,eberlein2008plasmon,abstreiter1984light}.
Fig.\,\ref{figure2}a shows the calculated $L(q,\omega)$ assuming typical experimental conditions:
$\mu=0.3\,$eV, $\Delta=0\,$eV, $T=300\,$K, $\kappa=2.5$, and $\eta=10\,$meV.
The single particle continuums are also indicated: (1) intraband, (2) electron-hole interband and
(3) low-energy interband.
The $\sqrt{q}$-plasmon lies above the intraband continuum,
and compares well with the long wavelength dispersion $\omega_{pl}(q)$,
while the $\gamma$-plasmon is significantly broadened.
The most important result is the appearance of distinctively sharp structure
near $\omega\approx\omega_{op}$, not seen in monolayer graphene\cite{WSSG06,HSS07}.

Fig.\,\ref{figure2}b plots the loss spectra at different momenta $q$.
We observed an enhancement in the IR activity of the phonon mode
as the plasmon resonance approaches $\omega_{op}$.
\textcolor{black}{The transfer of plasmonic spectral weight to the IR phonon mode,
as reflected by an increase in both intensity and linewidth,
enhances with decreasing detuning.
}Renormalized by many-body interactions, this `dressed' phonon exhibits
pronounced IR activity, and is also accompanied by a Fano asymmetric
spectral line-shapes.
The Fano feature is acquired through interference between the discrete phonon
mode and the `leaky' plasmonic mode; the electronic
lifetime is significantly shorter than that of the phonon, broadening
the former into a quasi-continuum.
The loss spectra show the evolution of the plasmonic and phonon resonances
as they approach each other. They evolve from separate resonances
at small $q$ to a Fano line-shape, and eventually an induced narrow
transparency at zero detuning.
This very narrow transparent window emerged within the
broadly opaque plasmonic absorption, a phenomenon analogous to the electromagnetically-induced transparency\cite{luk2010fano},
and should also be accompanied by novel electromagnetic effects such as slow light\cite{sandtke2007slow}.
\textcolor{black}{On the contrary, plasmon coupling with substrate surface optical phonons typically leads to well-separated resonances
instead\cite{YLZW13,hwang2010plasmon}.
}

Transmission spectroscopy studies has proven to be very effective in probing the plasmonic properties of graphene,
where finite plasmon momentum $q$ can be sampled by simply patterning graphene into
nanostructures\cite{JGHGM11,YLC12}. Graphene nanostructures with dimensions down to
$100\,$nm would allow us to access these predicted mid-IR plasmonic features
under experimentally accessible doping conditions\cite{YLC12}.
The enhancement of IR phonon activity with
decreased detuning between the phonon and plasmon resonance 
might lead to interesting applications.
Indeed, such plasmon-enhanced IR absorption
has permitted an emerging field of spectroscopy by noble metals of surfaces and electrochemical systems\cite{aroca2004surface}.
Tunable plasmonic resonance in graphene nanostructured surfaces might allow
for detection of molecules through enhancement of its IR vibrational modes.

\begin{figure}[t]
\centering
\scalebox{0.45}[0.45]{\includegraphics*[viewport=90 80 550 730]{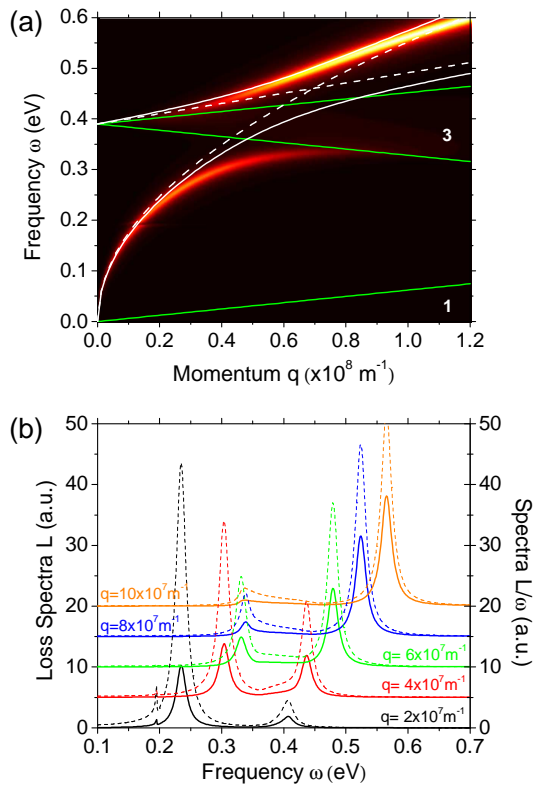}}
\caption{
$\bold{(a)}$ shows the RPA electron loss function $L(q,\omega)$ for bilayer graphene
computed at $T=300\,$K
at chemical potential of $\mu=0.6\,$eV, constant damping of $\eta=10\,$meV,
zero energy gap (i.e. $\Delta=0\,$eV) and a background dielectric constant of $\kappa=1$.
Spectra $L$ (solid lines) and $L/\omega$ (dashed lines) at different plasmon momenta $q$ are plotted in $\bold{(b)}$.
}
\label{figure3}
\end{figure}

Previously, we have seen that the $\gamma$-plasmon mode is overdamped.
In the limit of small momenta, it has the following dispersion\cite{G11},
\begin{align}
\omega_{\gamma}(q)=\frac{1}{\hbar}\left[\gamma + \frac{qe^2}{8\pi\epsilon_0\kappa}\mbox{log}\left(1+2\frac{\mu}{\gamma}\right)\right].
\label{om_gam}
\end{align}
If the $\gamma$-plasmon gains sufficient oscillator strength, e.g.
by modifying its doping  ($\uparrow$$\mu$) or dielectric environment ($\downarrow$$\kappa$),
it can reside outside the low-energy interband continuum.
This is shown in Fig.\,\ref{figure3}a (dashed line),
calculated using
Eq.\,\ref{om_gam} assuming $\mu=0.6\,$eV and  $\kappa=1$.
The electron loss function in Fig.\,\ref{figure3}a indicates
several interesting features of this high energy  $\gamma$-plasmon mode.
First, its dispersion departs from the simple $\omega_{\gamma}-\gamma\propto q$ relation,
acquiring an increasingly $q^2$ behavior with $q$.
We find that the modified dispersion can be described
within a model that accounts for the effective coupling between
the classical and $\gamma$-plasmon as follows,
\begin{align}
\epsilon_{eff}\approx\kappa\left[1-\frac{\omega^2_{pl}}{\omega^2}-\frac{\alpha^2}{\omega^2-\omega^2_{\gamma}+\alpha^2}\right],
\end{align}
where $\alpha$ is an effective coupling between the two modes.
Using the long-wavelength expressions for these modes, i.e. Eq.\,\ref{om_pl} and \ref{om_gam}
(dashed white lines), and
a coupling energy $\alpha=85\,$meV,
the coupled mode solutions (solid white lines) obtained by solving for $\epsilon_{eff}=0$
agrees well with the dispersions observed in the loss function.
Second, we observed prominent spectral weight transfer from
the conventional 2D plasmon to the $\gamma$-plasmon mode.

Fig.\,\ref{figure3}b plots the calculated $L(q,\omega)$ and $L(q,\omega)/\omega$ spectra
at typical values of $q=2-10\times 10^{7}\,$m$^{-1}$.
The integrated loss function $\int_{0}^{\infty}  L(q,\omega) d\omega$
is related to the Coulomb energy stored in the electron fluid\cite{pines1966theory}.
On the other hand, through the Kramers-Kronig relations, one can obtain the
sum rule $\int_{0}^{\infty}  L(q,\omega)/\omega d\omega=-1/\pi$\cite{marelarxiv},
with conserved spectral weight at different $q$.
We see that the $\gamma$-plasmon acquires a spectral weight an order larger
than the conventional plasmon as the latter enters into the Landau damped region.
Hence, it should be experimentally observable.
The possibility of
an `optical'-like high energy plasmonic mode,
previously presumed to be
overdamped with little spectral weight\cite{G11},
might open up applications in higher mid-IR spectral range.
With high enough doping, e.g. with electrolyte gating, this mode
can gain enough oscillator strength and be pushed out of the Landau
damped region, to become a coherent plasmonic mode.

In summary, we have shown that bilayer graphene as a new plasmonic material,
is important and interesting in its own right.
The above-mentioned new mid-IR plasmonic effects can also be generalized to more complex graphene stacks\cite{GNP06}, 
\textcolor{black}{for example ABC or ABA trilayers.
These new plasmonic resonant features can also potentially lead to interesting applications such as 
engineered metamaterials with novel electromagnetic effects\cite{YTGAX13},
resonant heat transfer processes\cite{shen2009surface}, among many others\cite{luk2010fano}.}

\emph{Acknowledgement:} FG acknowledges financial support from the Spanish Ministry of Economy (MINECO) through Grant no. FIS2011-23713, from the European Research Council Advanced Grant, contract 290846, and from European Commission under the Graphene Flagship contract CNECT-ICT-604391.


\end{document}